\documentclass[useAMS,usenatbib]{mn2e}

\usepackage{graphicx}

\renewcommand{\ln}{\mathrm{ln}}
\newcommand{\const}{\mathrm{const}}
\renewcommand{\cos}{\mathrm{cos}}
\newcommand{\Min}{\mathrm{Min}}

\title[On the effective particle acceleration in the paraboloidal magnetic field]
{On the effective particle acceleration in the paraboloidal
magnetic field}
\author[V. S. Beskin and E. E. Nokhrina]{V. S. Beskin$^{1}$\thanks{E-mail:
beskin@lpi.ru} and E. E. Nokhrina$^{2}$
\\
$^{1}$P.N.Lebedev Physical Institute, Leninsky prosp., 53, Moscow, 119991, Russia\\
$^{2}$Moscow Institute of Physics and Technology, Dolgoprudny,
Moscow region, 141700, Russia}
\begin{document}

\date{Accepted 1988 December 15. Received 1988 December 14; in original form 1988 October 11}

\pagerange{\pageref{firstpage}--\pageref{lastpage}} \pubyear{2002}

\maketitle

\label{firstpage}

\begin{abstract}
The problem of the efficiency of particle acceleration for
paraboloidal poloidal magnetic field is considered within the
approach of steady axisymmetric MHD flow. For the large Michel
magnetization parameter $\sigma$ it is possible to linearize the
stream equation near the force-free solution and to solve the
problem self-consistently as was done by \citet{BKR} for monopole
magnetic field. It is shown that on the fast magnetosonic surface
the particle Lorentz factor $\gamma$ does not exceed the standard
value $\sigma^{1/3}$. On the other hand,
in the supersonic region the Lorentz factor grows with the
distance $z$ from the equatorial plane as $\gamma \approx
(z/R_{\rm L})^{1/2}$
up to the distance $z \approx\sigma^2 R_{\rm L}$, where $R_{\rm
L}=c/\Omega_{\rm F}$ is the radius of the light cylinder. Thus,
the maximal Lorentz factor is $\gamma_{\rm max}\approx\sigma$,
which corresponds to almost the full conversion of the Poynting
energy flux into the particle kinetic one.
\end{abstract}

\begin{keywords}
MHD -- stream equation -- relativistic jets -- galaxies:jets
\end{keywords}

\section{Introduction}

An activity of many astrophysical sources (pulsars, active
galactic nuclei) is associated with the presence of a strong
magnetic field ($\sim 10^{4}$ G for AGNs and $\sim 10^{12}$ G for
pulsars) surrounding the rapidly rotating object and a
relativistic particle outflow. Convenient way to characterize such
flows is to introduce the magnetization parameter $\sigma$, as
was first done by \citet{Mchl}:
\begin{equation}
\sigma=\frac{e\Omega\Psi_{\rm tot}}{4\lambda m_{\rm e} c^{3}}.
\label{sigma}
\end{equation}
Here $\Psi_{\rm tot}$ is the total magnetic flux, and $\lambda =
n/n_{\rm GJ}$ is the multiplication parameter of plasma ($|e|
n_{\rm GJ} = {|\mbox{\boldmath$\Omega$}}\textbf{B}|/2\pi c$ is
the Goldreich-Julian charge density). The magnetization parameter
characterizes the quotient of electro-magnetic flux to particle
kinetic energy flux near the surface of an object. The so-called
$\sigma$-problem, i.e., the problem of transformation of
electro-magnetic energy into particle kinetic one, appears while
one trying to explain an effective particle acceleration in the
magnetic field.

Indeed, the flow in the vicinity of different objects is assumed
to be strongly magnetized at its origin. In spite of the lack of
observational data near the surface of radio pulsars, theoretical
modeling predicts that the wind in this region has a composition
$\sigma\gg 1$ \citep{michbook,lsurf1}. The same can be said about
the AGNs \citep{BBR}. As for the Lorentz factor $\gamma$, at
least for the blazars, the deficiency of a soft X-ray bump in
their spectrum, which would be produced by the Comptonized direct
radiation from the disc, is a reason to exclude the presence of
particles with $\gamma>5$ in the near zone of an object
\citep{SBML}. On the other hand, at large distance from a pulsar,
observations and modeling allow us to determine the magnetization
parameter $\sigma\approx 10^{-3}$ \citep{KC}. Observations of
quasars and active galactic nuclei give $\sigma\leq 1$ and
$\sigma\ll 1$ \citep{SBML}.

Up to now the axisymmetric stationary MHD approach gave the
inefficient particle acceleration beyond the fast magnetosonic
surface \citep{BKR,bgvl2,acc}. This seemed to be a general
conclusion for any structure of a flow, however a lack of
acceleration in the supersonic region was rather the consequence
of monopole field \citep{SpitArons, Thom}. The reconnection in a
striped wind was discussed as one of possible acceleration
scenarios \citep{recon1,recon2,recon3}; however, this mechanism
can account only for the non-axisymmetric part of a Poynting flux.
Besides, for the Crab pulsar, \citet{recon2} had shown that
effective acceleration may take place only beyond the standing
shock. Another acceleration process can be connected with possible
restriction of the longitudinal current, and thus the appearance
of a light surface $|\mathbf{E}|=|\mathbf{B}|$ at the finite
distance from a central object. In this case the effective energy
conversion and the current closure takes place in the boundary
layer near the light surface \citep{lsurf1,lsurf2,lsurf3}.

In this work we are trying to solve the problem of particle
acceleration self-consistently within the approach of stationary
axisymmetric magnetohydrodynamics. We regard particle inertia to
be a small disturbance to the force-free flow. This allows us to
linearize stream equation and to find the disturbance to the
magnetic flux, which corresponds to the finite mass of particles.
Given that we can find the growth of a Lorentz factor. In the work
of \citet{BKR} the Michel's monopole solution was taken as the
zero approximation to the flow. For this structure of the magnetic
field, the Lorentz factor was found to be $\sigma^{1/3}$ on the
fast magnetosonic surface, which was located at the finite
distance unlike the force-free limit. Beyond that singular
surface the acceleration turned out to be ineffective. Treating
the problem numerically, \citet{bgvl2} and later \cite{Komm} also
has got inefficient acceleration and collimation for monopole
outflow. Now we took the flow near another force-free solution,
i.e., Blandford's paraboloidal magnetic field \citep{Bland}. The
obvious difference of this zero-approximation in comparison with
monopole solution is a well-collimated flow even in the
force-free limit.

In Section 2 we remind the trans-field and Bernoulli equations
describing the stationary axisymmetric ideal outflow. After
formulating the problem in Section 3, it is shown in Section 4
that the fast magnetosonic surface (FMS) is located at the
distance $r_{\rm F}=R_{\rm L} (\sigma/\theta)^{1/2}$ from the
central object, the Lorentz factor $\gamma$ being
$(\sigma\theta)^{1/2}$ on it. This result is consistent with the
standard value $\gamma \leq \sigma^{1/3}$. In Section 5 it is
shown that the disturbance of magnetic flux due to the finite
mass of particles remains small in the subsonic region, the
Lorentz factor growing linearly with a distance from the
rotational axis within the FMS. In Section 6 we show that the
supersonic flow in the paraboloidal geometry is in fact
one-dimensional, and it is possible to treat the problem
numerically. In the supersonic region the growth of Lorentz
factor remains linear with the distance $\varpi$ from the
rotational axis, reaching the value $\sigma$ at the distance $z
\sim \sigma^2 R_{\rm L}$ from the equatorial plane. This
corresponds to almost the full conversion of the Poynting flux
into the particle kinetic energy flux. Finally in Section 7 we
discuss some astrophysical applications.

\section{Basic equations}

Let us consider a stationary axisymmetric MHD flow of cold plasma
in a flat space. Within this approach magnetic field is expressed
by
\begin{equation}
\textbf{B}=\frac{\nabla\Psi\times\textbf{e}_{\hat{\varphi}}}{2\pi\varpi}-\frac{2
I}{\varpi}\; \textbf{e}_{\hat{\varphi}}.
\end{equation}
Here $\Psi(r,\theta)$ is the stream function, $I(r,\theta)$ is
the total electric current inside magnetic tube
$\Psi(r,\theta)=\const$, and $\varpi = r \sin\theta$ is the
distance from the rotational axis. In this paper we put $c=1$.
Owing to the condition of zero longitudinal electric field, one
can write down the electric field as
\begin{equation}
\textbf{E}=-\frac{\Omega_{\rm F}}{2\pi}\nabla\Psi,
\end{equation}
where the angular velocity $\Omega_{\rm F}$ is constant on the
magnetic surfaces: $\Omega_{\rm F}=\Omega_{\rm F}(\Psi)$. The
frozen-in condition $\textbf{E}+\textbf{v}\times\textbf{B}=0$
gives us
\begin{equation}
\textbf{u}=\frac{\eta}{n}\textbf{B}+\gamma\Omega_{\rm
F}\varpi\textbf{e}_{\hat{\varphi}}, \label{ufour}
\end{equation}
where $\textbf{u}$ is four-velocity of a flow, and $n$ is the
concentration in the comoving reference frame. Function $\eta$ is
the ratio of particle flux to the magnetic field flux. Using the
continuity equation $\nabla(n\textbf{u})=0$, one gets that $\eta$
is constant on magnetic surfaces as well: $\eta=\eta(\Psi)$.

Two extra integrals of motion are the energy flux, conserved due
to stationarity,
\begin{equation}
E(\Psi)=\frac{\Omega_{\rm F}I}{2\pi}+\mu\eta\gamma,
\end{equation}
and the $z$-component of the angular momentum, conserved due to
axial symmetry,
\begin{equation}
L(\Psi)=\frac{I}{2\pi}+\mu\eta\varpi u_{\hat{\varphi}}.
\end{equation}
Here $\mu$ is the relativistic enthalpy, which is a constant for
the cold flow. The fifth integral of motion is the entropy
$s(\Psi)$, which is equal to zero for the cold flow under
consideration.

If the flux function $\Psi$ and the integrals of motion are
given, all other physical parameters of the flow can be
determined using the following algebraic relations
\citep{Camenzind,Beskin97}:
\begin{equation}
\frac{I}{2\pi}=\frac{L-\Omega_{\rm F}\varpi^{2}E}{1-\Omega_{\rm
F}^{2}\varpi^{2}-{\cal M}^{2}},\label{I}
\end{equation}
\begin{equation}
\gamma=\frac{1}{\mu\eta}\cdot\frac{E-\Omega_{\rm F}L-{\cal
M}^{2}E}{1-\Omega_{\rm F}^{2}\varpi^{2}-{\cal
M}^{2}},\label{gamma}
\end{equation}
\begin{equation}
u_{\hat{\varphi}}=\frac{1}{\varpi\mu\eta}\cdot\frac{(E-\Omega_{\rm
F}L)\Omega_{\rm F}\varpi^{2}-{\cal M}^{2}L} {1-\Omega_{\rm
F}^{2}\varpi^{2}-{\cal M}^{2}},\label{u}
\end{equation}
where the Alfv\'enic Mach number ${\cal M}$ is
\begin{equation}
{\cal M}^{2}=\frac{4\pi\eta^{2}\mu}{n}.
\end{equation}
To determine  ${\cal M}^{2}$, one should use the definition of
Lorentz factor $\gamma^{2}-\textbf{u}^{2}=1$ which gives the
Bernoulli equation in the form
\begin{equation}
\frac{K}{\varpi^{2}A^{2}}=\frac{1}{64\pi^{4}}\cdot\frac{{\cal
M}^{4}
(\nabla\Psi)^{2}}{\varpi^{2}}+\mu^{2}\eta^{2}.\label{Bern-0}
\end{equation}
Here
\[
A=1-\Omega_{\rm F}^{2}\varpi^{2}-{\cal M}^{2},
\]
\[
K=\varpi^{2}(E-\Omega_{\rm F} L)^{2}(1-\Omega_{\rm
F}^{2}\varpi^{2}-2{\cal M}^{2})+{\cal
M}^{4}(\varpi^{2}E^{2}-L^{2}).
\]

The cold transonic flow is characterized by two singular
surfaces: the Alfv\'enic surface and the fast magnetosonic surface
(FMS). The first is determined by the condition of nulling the
denominator $A$ in the relations (\ref{I})--(\ref{u}). FMS can be
defined as the singularity of Mach number's gradient. Writing
equation (\ref{Bern-0}) in the form
\[
(\nabla\Psi)^{2}=F=\frac{64\pi^{4}}{{\cal
M}^{4}}\cdot\frac{K}{A^{2}}- \frac{64\pi^{4}}{{\cal
M}^{4}}\varpi^{2}\mu^{2}\eta^{2},
\]
and taking the gradient of it, we can get
\[\nabla_{k}{\cal M}^{2}=\frac{N_{k}}{D}.
\]
Here
\[D=\frac{A}{{\cal M}^{2}}+\frac{1}{{\cal M}^{2}}\cdot\frac{B_{\varphi}^{2}}{B_{\rm P}^{2}},
\]
\[N_{k}=-\frac{A\nabla^{i}\Psi\cdot\nabla_{k}\nabla_{i}\Psi}
{(\nabla\Psi)^{2}}+\frac{A}{2(\nabla\Psi)^{2}}\nabla_{k}^{'}F,
\]
and the operator $\nabla_{k}^{'}$ acts on all quantities except
for ${\cal M}^{2}$. The regularity conditions
\[D=0;\; N_{r}=0;\;
N_{\theta}=0
\]
define the position of FMS and the relation between the integrals
of motion on it.

Finally, the stream equation on the function $\Psi(r,\theta)$ can
be written as \citep{Beskin97}
\begin{equation}
\begin{array}{l}
\displaystyle\nabla_{k}\left[\frac{1}{\varpi^{2}}\left(1-\Omega_{\rm
F}^{2}
\varpi^{2}-{\cal M}^{2}\right)\nabla^{k}\Psi\right]+\\\ \\
\displaystyle +\Omega_{\rm F}(\nabla\Psi)^{2}\frac{d\Omega_{\rm
F}}{d\Psi}+\frac{64\pi^{4}}{\varpi^{2}}\frac{1}{2{\cal M}^{2}}
\frac{\partial}{\partial\Psi}\left(\frac{G}{A}\right)=0,
\label{GS-0}
\end{array}
\end{equation}
where
\begin{equation}
G=\varpi^{2}(E-\Omega_{\rm F}L)^{2}+{\cal M}^{2}L^{2}-{\cal
M}^{2}\varpi^{2}E^{2}.
\end{equation}
Operator $\partial/\partial\Psi$ acts only on the integrals of
motion. The stream equation (\ref{GS-0}) contains the magnetic
flux function $\Psi$ and four integrals of motion: $E(\Psi)$,
$L(\Psi)$, $\eta(\Psi)$, and $\Omega_{\rm F}(\Psi)$, i.e., it has
the Grad-Shafranov form.

\section{The problem}

Our goal is to determine the characteristics of a flow in the
paraboloidal magnetic field. For this reason it is convenient to
use the following orthogonal coordinates:
\[
X=r(1-\cos{\theta});\; Y=r(1+\cos{\theta});\; \varphi.
\]
Here $X$ stands for the certain magnetic surface in the force-free
Blandford's paraboloidal solution, $Y$ is the distance along it,
and $\varphi$ is the azimuthal angle. The latter does not appear
in the equations because of the axial symmetry of the problem. The
flat metric in this coordinates is
\[g_{\rm XX}=\frac{X+Y}{4X};\; g_{\rm YY}=\frac{X+Y}{4Y};\;
g_{\varphi\varphi}=XY.\]

Then, Blandford's force-free solution can be written down as
\citep{LeePark}:
\begin{equation}
\frac{d\Psi}{d X}=\frac{\pi{\cal C}}{\sqrt{1+\Omega_{\rm
F}^{2}(X)X^{2}}},
\end{equation}
\begin{equation}
I_{0}(\Psi)=\frac{{\cal C}\Omega_{\rm
F}(X)X}{2\sqrt{1+\Omega_{\rm F}^{2}X^{2}}},
\end{equation}
where $\Omega_{\rm F}$ is the arbitrary function of $\Psi$, and
$\cal C$ is a constant. In particular, for $\Omega_{\rm F}=\const$
we have
\begin{equation}
\Psi_{0}(X)=\frac{\pi {\cal C}}{\Omega_{\rm
F}}\ln\left(\Omega_{\rm F}X+\sqrt{(\Omega_{\rm F}X)^{2}+1}\right).
\label{exact}
\end{equation}

\begin{figure}
\includegraphics{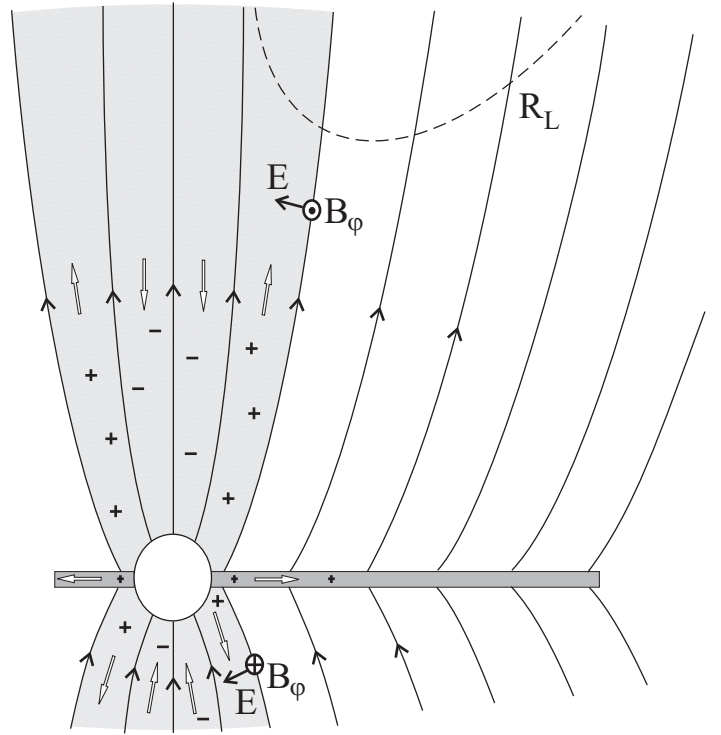}
\caption{Blandford's force-free paraboloidal solution. Here the
thick arrows represent the volume currents. The shadowed region
is the working volume $\Omega_{\rm F} \approx$ const. The dashed
line represents the light cylinder.}
\end{figure}

We assume that in the vicinity of a central object the particle
energy flux is much smaller than that of electro-magnetic field.
In this case it is possible to consider the contribution of
particle inertia as a small disturbance to the quantities of the
force-free flow. Thus, in the first approximation we can get from
(\ref{GS-0}) the linear equation on the disturbance and solve the
problem self-consistently.

As was already stressed, for cold plasma the problem is
characterized by two singular surfaces: Alfv\'enic and fast
magnetosonic ones. Consequently, we need to specify four boundary
conditions on the disc surface $D$ \citep{Beskin97}. For
simplicity, we consider the case
\[\Omega_{\rm F}(\Psi)|_{\rm D}=\const=\Omega_{\rm F},\]
\[\gamma|_{\rm D}=\const=\gamma_{\rm in}\ll\sigma^{1/3},\]
\[\eta(\Psi)|_{\rm D}=\const=\eta,\]
\[\Psi|_{\rm D}=\frac{\pi {\cal C}}{\Omega_{\rm F}}\ln\left(\Omega_{\rm F}\varpi+
\sqrt{(\Omega_{\rm F}\varpi)^{2}+1}\right).\] The condition
$\Omega_{\rm F}=\const$ naturally restricts the region of a flow
under consideration (see Fig.~1). Since we assume that the
magnetic field is frozen in the disc, we must consider the flow
only when $\Omega_{\rm F}X|_{D}=\Omega_{\rm F}\varpi<1$. In fact,
we should use the inequality $\Omega_{\rm F}X\ll 1$. Then,
Michel's magnetization parameter $\sigma$ for our problem can be
defined as
\begin{equation}
\sigma=\frac{E_{\rm A}}{\mu\eta}=\frac{{\cal C}\Omega_{\rm
F}}{8\pi\mu\eta}.\label{sigma-par} \label{sigmadef}
\end{equation}
Here $E_{\rm A}$ is a kind of energy amplitude:
\begin{equation}
E=\frac{{\cal C}\Omega_{\rm F}^{2}X}{4\pi\sqrt{1+\Omega_{\rm
F}^{2}X^{2}}}\approx\frac{{\cal C}\Omega_{\rm
F}^{2}X}{4\pi}=E_{\rm A}\cdot 2 \Omega_{\rm F}X.
\end{equation}
Under our assumptions $\sigma\gg 1$, and we can introduce the
small quantity $\varepsilon=\sigma^{-1}$. Besides, we will be
mainly interested in the flow far from the light cylinder, and so
we have another limitation $\Omega_{\rm F}^{2}XY\gg 1$.

We shall seek the stream function of the problem in the form
\begin{equation}
\Psi(X,Y)=\Psi_{0}(X)+\varepsilon f(X,Y),
\end{equation}
where $\varepsilon f(X,Y)$ is the disturbance to find. The
function of the angular momentum $L$ may in general be different
from the function $L_{0}$. For $L$ the following expression may
be written:
\begin{equation}
L(\Psi)=L(\Psi_{0})+\varepsilon f\frac{\partial
L(\Psi_{0})}{\partial\Psi_{0}}=L(\Psi_{0})+\varepsilon l.
\end{equation}

\section{Fast magnetosonic surface}

In order to find the position of FMS one can rewrite the Bernoulli
equation (\ref{Bern-0}) in the form
\begin{equation}
\begin{array}{l}
\displaystyle q^{4}+2q^{3}-\left(\xi+\frac{1}{\Omega_{\rm
F}^{2}\varpi^{2}}\right)q^{2}+
2q\left(\frac{\mu^{2}\eta^{2}}{E^{2}}+\frac{(e')^{2}}{\Omega_{\rm F}\varpi^{2}}\right)+\\\ \\
\displaystyle
+\frac{\mu^{2}\eta^{2}}{E^{2}}+\frac{(e')^{2}}{\Omega_{\rm
F}\varpi^{2}}=0.\label{Bernoulli-2}
\end{array}
\end{equation}
Here by definition $q={\cal M}^{2}/\Omega_{\rm F}^{2}\varpi^{2}$,
\begin{equation}
\xi=1-\frac{\Omega_{\rm
F}^{4}\varpi^{2}(\nabla\Psi)^{2}}{64\pi^{4}E^{2}},
\end{equation}
and $e'=E(\Psi)-\Omega_{\rm F}L(\Psi)=\mu\eta\gamma(1-\Omega_{\rm
F}\varpi v_{\hat{\varphi}})\approx\gamma_{\rm in}\mu\eta=\const$.
It is easy to check that for the force-free solution
\begin{equation}
\displaystyle\xi=\frac{1-\cos{\theta}}{2}\approx\frac{\theta^{2}}{4}\ll
1
\end{equation}
for the small angle $\theta$, i.e., in the whole region near the
rotational axis. Remember that $\xi\equiv 0$ for Michel
force-free monopole outflow.

To show that the quantity $q$ is much smaller than unity, one can
write (\ref{gamma}) assuming $\sigma\gg 1$ and $\Omega_{\rm
F}^{2}\varpi^{2}\gg 1$. Solving this equation for $q$, we get
\begin{equation}
q=\frac{\gamma\mu\eta}{E} \approx \frac{\gamma}{\sigma}.
\end{equation}
Thus, $q$ is approximately equal to the ratio of particle kinetic
energy to the full energy of the flow, i.e., $q\ll 1$ for the
magnetically dominated flow.

As a result, one can rewrite (\ref{Bernoulli-2}) in the form
\begin{equation}
2q^{3}-\left(\xi+\frac{1}{\Omega_{\rm
F}^{2}\varpi^{2}}\right)q^{2}+
\frac{\mu^{2}\eta^{2}}{E^{2}}+\frac{(e')^{2}}{\Omega_{\rm
F}\varpi^{2}}=0,\label{Bernoulli-3}
\end{equation}
where the terms $q^{4}$ and $q$ were omitted due to their
smallness.

Fast magnetosonic surface corresponds to the intersection of two
roots of equation (\ref{Bernoulli-3}), or to the condition of
discriminant $Q$ being equal to zero \citep{BKR}. The regularity
conditions give $\partial Q/\partial r=0$ and $\partial
Q/\partial\theta=0$, or $\partial Q/\partial X=0$ and $\partial
Q/\partial Y=0$. For equation (\ref{Bernoulli-3}) discriminant
$Q$ is expressed by
\begin{equation}
Q=\frac{1}{16}\frac{\mu^{4}\eta^{4}}{E^{4}}-
\frac{1}{16}\frac{1}{27}\frac{\mu^{2}\eta^{2}}{E^{2}}\left(\xi+\frac{1}{\Omega_{\rm
F}^{2}\varpi^{2}}\right)^{3}.
\end{equation}
The condition of root intersection $Q=0$ can be rewritten as
\begin{equation}
\xi(r_{\rm F},\theta)+\frac{1}{\Omega_{\rm F}^{2}r_{\rm
F}^{2}\theta^{2}}=3\left(\frac{\mu\eta}{E}\right)^{2/3},
\end{equation}
and the first regularity condition $\partial Q/\partial Y$ as
\begin{equation}
Y\frac{\partial\xi}{\partial Y}=\frac{1}{\Omega_{\rm F}^{2}XY}.
\label{xi-fms}
\end{equation}
Taking approximately $\xi\approx Y\cdot\partial\xi/\partial Y$ we
get for the position of FMS
\begin{equation}
r_{\rm F}(\theta) \approx R_{\rm
L}\left(\frac{\sigma}{\theta}\right)^{1/2},\label{fms}
\end{equation}
where $R_{\rm L}=\Omega_{\rm F}^{-1}$ is the radius of a light
cylinder.

\begin{figure}
\includegraphics{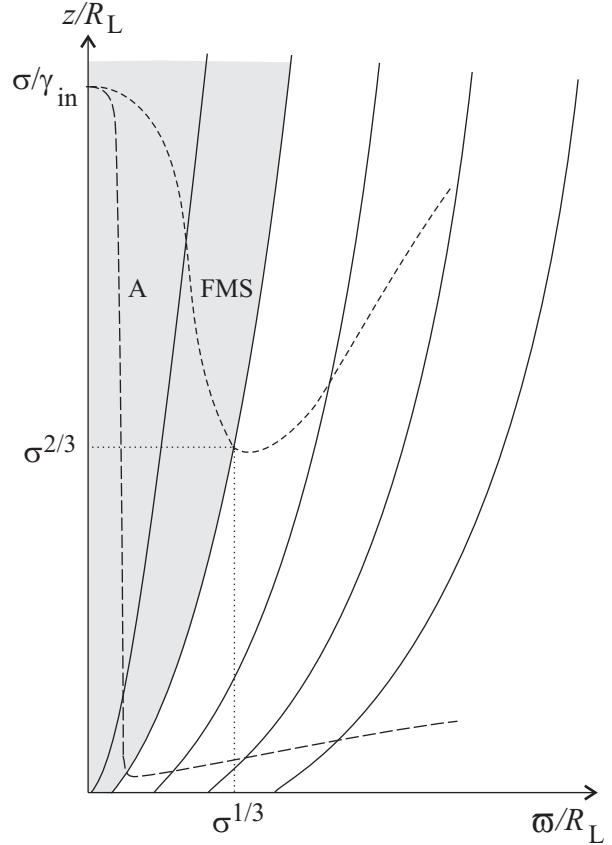}
\caption{Position of the Alfv\'enic surface (long dashed line) and
FMS (short dashed lines) for paraboloidal magnetic field. The
self-similar flow can be realized only outside the working
volume.}
\end{figure}

Values of $q(r_{\rm F},\theta)$ and $\gamma_{\rm F} =
\gamma(r_{\rm F},\theta)$ due to condition $Q=0$ do not depend on
the sum $(\xi+1/\Omega_{\rm F}^{2}\varpi^{2})$ and on the FMS are
equal to
\begin{equation}
q(r_{\rm F},\theta)=\frac{1}{\sigma\theta},
\end{equation}
\begin{equation}
\gamma_{\rm F}=(\sigma\theta)^{1/2}.
\end{equation}
Again we confirm that $q\ll 1$ since $\sigma\theta=\Omega_{\rm
F}^{2}XY|_{r_{\rm F}}\gg 1$. These results are valid when
\begin{equation}
\left(\frac{\mu\eta}{E}\right)^{2}\gg\left(\frac{e'}{E\Omega_{\rm
F}\varpi}\right)^{2},
\end{equation}
i.e., in the region where electro-magnetic energy is greater than
kinetic one. On the FMS this region is defined by the angle
$\theta$ changing from $\gamma_{\rm in}^{2}\sigma^{-1}$ to
$\sigma^{1/3}$. Here the greatest value of $\theta$ is given by
the condition $\Omega_{\rm F}X=1$, which corresponds to the
boundary of the working volume. On the other hand, for
$\theta<\gamma_{\rm in}^{2}\sigma^{-1}$ one can get
\begin{equation}
r_{\rm F} \approx R_{\rm L}\frac{\sigma}{\gamma_{\rm
in}},\label{fms-axis}
\end{equation}
\begin{equation}
\gamma_{\rm F}=\gamma_{\rm in}.
\end{equation}
As we see, along the rotational axis particle energy remains the
same as near the origin.

The position of the FMS on the rotational axis can be evaluated
independently. Indeed, for $B_{\varphi}=0$ the condition $D=0$
coincides with the condition $A=0$, i.e., the position of the
fast magnetosonic surface coincides with the Alfv\'enic surface
on the axis. Assuming that $u_{\rm p}\approx\gamma_{\rm in}$ and
using the definitions (\ref{ufour}) and (\ref{sigmadef}) one can
obtain
\begin{equation}
{\cal M}^2 = \frac{\sigma}{\gamma_{\rm in}} \,  \Omega_{\rm F} r.
\end{equation}
It gives
\begin{equation}
r_{\rm F} \approx r_{\rm A}\approx \frac{\sigma}{\gamma_{\rm in}}
R_{\rm L},
\end{equation}
coinciding with (\ref{fms-axis}). This distance is much larger
than the appropriate radius $r_{\rm F} \approx
(\sigma/\gamma_{\rm in})^{1/2} R_{\rm L}$ for monopole magnetic
field. But the transverse dimension of the working volume $\varpi
\approx \sigma^{1/3} R_{\rm L}$ remains the same. We see as well
that on the FMS within the working volume the value of Lorentz
factor changes from $\gamma_{\rm in}$ to $\sigma^{1/3}$. The
greatest value of $\gamma$ is equal to the corresponding value
for the monopole structure \citep{BKR}.

Finally, as one can see in Fig.~2, the shape of the FMS extends
along the rotational axis. Thus, the working volume cannot be
described within the the self-similar approach as in this case
the singular surfaces are to have conical shape $\theta=\const$
\citep{BPayne, lcb92}.

\section{Subsonic flow}

For the inner region of the flow we shall write down the stream
equation with the small disturbances to the functions $\Psi$,
$L$, and $E$. We shall treat the quantities $q$ and $\varepsilon
f$ as being of the same order of smallness. In the zero
approximation one can get the equation
\begin{equation}
\begin{array}{l}
\displaystyle -\frac{4\Omega_{\rm
F}^{2}}{X+Y}\frac{\partial}{\partial
X}\left(X\frac{\partial\Psi_{0}}{\partial X}\right)+\\\ \\
\displaystyle
+\frac{64\pi^{4}}{XY}\frac{1}{\partial\Psi_{0}/\partial
X}L(\Psi_{0})\frac{\partial L(\Psi_{0})}{\partial X}=0.\label{1}
\end{array}
\end{equation}
Clearly, it has a solution (\ref{exact}) for $L(\Psi_0) =
L_0(\Psi_0)$. On the other hand, in the first approximation we
have
\begin{equation}
\begin{array}{llll}
\displaystyle -\frac{4\Omega_{\rm
F}^{2}}{X+Y}\left[q\frac{\partial}{\partial
X}\left(X\frac{\partial\Psi_{0}}{\partial
X}\right)+X\frac{\partial q}{\partial
X}\frac{\partial\Psi_{0}}{\partial X}\right]-\\\ \\
\displaystyle -\frac{4\Omega_{\rm
F}^{2}}{X+Y}\varepsilon\left(X\frac{\partial^{2}f}{\partial
X^{2}}+ \frac{\partial f}{\partial
X}+Y\frac{\partial^{2}f}{\partial
Y^{2}}+\frac{\partial f}{\partial Y}\right)+\\\ \\
\displaystyle
-\frac{1}{2}\cdot\frac{64\pi^{4}}{XY}q\frac{1}{\partial\Psi_{0}/\partial
X}\frac{\partial L^{2}(\Psi_{0})}{\partial X}-\\\ \\
\displaystyle
+\frac{1}{2}\cdot\frac{64\pi^{4}}{XY}\varepsilon\frac{\partial
f/\partial X}{(\partial\Psi_{0}/\partial X)^{2}}\frac{\partial
L^{2}(\Psi_{0})}{\partial
X}+\\\ \\
\displaystyle
+\frac{64\pi^{4}}{XY}\varepsilon\frac{1}{\partial\Psi_{0}/\partial
X}\frac{\partial (L(\Psi_{0})l)}{\partial X}+\\\ \\
\displaystyle
+\frac{64\pi^{4}}{XY}\varepsilon\frac{e'}{\Omega_{\rm
F}}\frac{1}{\partial\Psi_{0}/\partial X}\frac{\partial
L(\Psi_{0})}{\partial X}=0.\label{2}
\end{array}
\end{equation}
The integral $L$ depends on the variable $Y$ as
$\Psi=\Psi_{0}+\varepsilon f$ depends on it.

Before we proceed to solve equation (\ref{2}) we can evaluate the
ratio $\varepsilon f/\Psi_{0}$ on the fast magnetosonic surface.
In order to do this, we need to express $\xi$ as a function of
$\varepsilon f$:
\begin{equation}
\xi=\frac{2\varepsilon}{\pi {\cal
C}}\left(\frac{f}{X}-\frac{\partial f }{\partial X}\right).
\end{equation}
Given the order of $\xi$ on the FMS by (\ref{xi-fms}), one can get
\begin{equation}
\frac{\varepsilon f}{\pi {\cal C} X}\sim\frac{1}{\Omega_{\rm
F}^{2}X Y}.
\end{equation}
Thus,
\begin{equation}
\frac{\varepsilon f}{\Psi_{0}}\sim\frac{1}{\sigma\theta}\ll 1,
\end{equation}
where $\sigma\theta\gg 1$ as we are interested in the flow
structure outside the light cylinder. Hence, our disturbance
proved to be small in comparison with the force-free solution up
to the fast magnetosonic surface. Finally, from (\ref{2}) we can
get an equation on the function $\varepsilon f$:
\begin{equation}
\begin{array}{l}
\displaystyle \varepsilon X^2\frac{\partial^{2} f}{\partial
X^{2}}+\varepsilon X \frac{\partial f}{\partial X} -\varepsilon f
+\varepsilon XY\frac{\partial^{2} f}{\partial Y^{2}} +
\varepsilon X\frac{\partial f}{\partial
Y}\\\ \\
\displaystyle + \pi {\cal C}\left(2qX +X^2\frac{\partial
q}{\partial X}\right)+\frac{4\pi^{2}e'}{\Omega_{\rm F}^{2}}=0.
\label{r<rf}
\end{array}
\end{equation}

To obtain $q$, we shall make a natural assumption that $q$ and
$\xi$ grow monotonically from correspondingly $\sigma^{-1}$ and
$0$ near the origin of the flow to $\left(\mu\eta/E\right)^{2/3}$
and $1/\Omega_{\rm F}^{2}\varpi^{2}$ on the fast magnetosonic
surface. Thus, in (\ref{Bernoulli-3}) one can neglect the terms
$q^{3}$ and $\xi q^{2}$ in comparison with $(\mu\eta/E)^{2}$ and
$q^{2}/(\Omega_{\rm F}\varpi)^{2} $ correspondingly. After that
the solution of (\ref{Bernoulli-3}) is expressed by
\begin{equation}
q=\frac{1}{2\sigma}\left(\frac{Y}{X}\right)^{1/2}.
\end{equation}

It is necessary to stress that in equation (\ref{r<rf}) we can
neglect the derivatives over $Y$ in comparison with the
derivatives over $X$. Here we take into account our assumptions
that $Y\gg X$, so that $\partial f/\partial X\sim f/X$ by the
order of magnitude. Inside the working volume this means that we
can neglect the curvature of field lines in our problem. Indeed,
one can find that
\begin{equation}
k=\frac{1}{R_{\rm c}}=\frac{1}{Y}\left(\frac{X}{Y}\right)^{1/2}
+\frac{2 \varepsilon}{(X Y)^{1/2}}
                \left(\frac{\partial f}{\partial Y}+
                Y\frac{\partial^{2}f}{\partial Y^{2}}\right),
\end{equation}
formally writing the expression for curvature $k$ for the implicit
function $\Psi_{0}(X(x,y))+\varepsilon f(X(x,y),Y(x,y))=\const$.
Here the term $(X^{1/2}/Y^{3/2})$ corresponds to the inverse
curvature radius of the force-free magnetic surfaces. Thus, the
curvature term does not play any role in the force balance on the
magnetic surfaces in the paraboloidal magnetic field, which was
quite different for the monopole magnetic field. In the latter
case the curvature term played the leading role in the asymptotic
region \citep{BeskinOkamoto}.

After substitution of the found function $q(X,Y)$ into
(\ref{r<rf}), one can find for the disturbance of the stream
function
\begin{equation}
\varepsilon f=\frac{\pi {\cal C}}{\sigma}\Omega_{\rm F} (X
Y)^{1/2},
\end{equation}
with the Lorentz factor being equal to
\begin{equation}
\gamma=\Omega_{\rm F}(X Y)^{1/2} = \Omega_{\rm F} r \sin\theta.
\end{equation}
Thus, in the subsonic region the Lorentz factor grows linearly
with the distance from the axis and reaches on the fast
magnetosonic surface the value $(\sigma\theta)^{1/2}$, which
corresponds to the result found in the previous Section. In this
sense one can say that the solution in the inner region of the
flow can be continued to the fast magnetosonic surface.

\section{Supersonic flow}

In order to solve the problem in the supersonic region we need to
emphasize some features of the paraboloidal configuration of
magnetic field.
\begin{enumerate}
\item The character of the flow may change in the
vicinity of the singular surface.
\item It was shown above that for paraboloidal magnetic field
the curvature of field lines does not play any role in the force
balance inside the working volume $\Omega_{\rm F}X<1$. This
allows us to consider the flow as one-dimensional.
\item Positions of the fast magnetosonic surfaces in the paraboloidal
field and in the cylindrical field coincide.
\end{enumerate}
Let us clarify the last point. The position of the fast
magnetosonic surface for the magnetized cylindrical jet immersed
into the external magnetic field $B_{\rm ext}$ is given by the
relation \citep{Cyl}
\begin{equation}
\Omega_{\rm F} \varpi \sim \sigma\frac{B_{\rm ext}}{\Omega_{\rm
F}^{2}\Psi_{\rm jet}},
\end{equation}
where $\varpi \approx r \theta$ is the distance from the
rotational axis, and $\Psi_{\rm jet}$ is the total flux inside
the jet. On the other hand, according to our definition of
$\sigma$ (\ref{sigma-par}) for paraboloidal flow
\begin{equation}
E=2\mu\eta\sigma\Omega_{\rm F}X=\frac{\Omega_{\rm
F}^{2}\Psi}{4\pi^{2}}.
\end{equation}
As a result, we shall define the magnetic flux inside the region
$\Omega_{\rm F}X<1$ as $\Psi_{\rm jet}$, and thus $\cal C$ is
expressed by
\begin{equation}
{\cal C}=\frac{\Psi_{\rm jet}\Omega_{\rm F}}{\pi}.
\end{equation}
In this case the expressions for $\sigma$ for two flows coincide.
Taking poloidal paraboloidal field as the external field for the
one-dimensional flow
\begin{equation}
B_{\rm p}=\frac{{\cal C}}{2 z}, \label{Bp}
\end{equation}
one can get the position of the FMS
\begin{equation}
r_{\rm F}\sim R_{\rm L}\left(\frac{\sigma}{\theta}\right)^{1/2},
\end{equation}
which coincides with (\ref{fms}).

Hence, the flow becomes actually 1D in the vicinity of the FMS,
not to say about the supersonic region. For this reason we can
consider the supersonic flow as one-dimensional.
But unlike \citet{Cyl} we would use the paraboloidal magnetic
field $B_{\rm p}(z)$ (\ref{Bp}) outside the working volume as an
external one. It gives us the slow $z$-dependence of all the
values.

For the cylindrical flow the integrals of motion near the axis
are the same as the integrals of motion in our paraboloidal
problem:
\begin{equation}
L(\Psi)=\frac{\Omega_{\rm F}\Psi}{4\pi^{2}},
\end{equation}
\begin{equation}
\Omega_{\rm F}(\Psi)=\const,
\end{equation}
\begin{equation}
\eta(\Psi)=\const,
\end{equation}
\begin{equation}
E(\Psi)=\gamma_{\rm in}\mu\eta+\Omega_{\rm F}L=e'+\Omega_{\rm F}L,
\end{equation}
where $e'=\const$. Introducing non-dimensional variables
\begin{equation}
y=\sigma\frac{\Psi}{\Psi_{0}},
\end{equation}
\begin{equation}
x=\Omega_{\rm F}\varpi,
\end{equation}
one can rewrite equations (\ref{Bern-0})--(\ref{GS-0}) as a set of
ordinary differential equations for $y$ and $M^{2}$ \citep{Cyl}:
\begin{equation}
\begin{array}{l}
\displaystyle (1-x^{2}-{\cal M}^{2})^{2}\left(\frac{d y}{d
x}\right)^{2}= \frac{\gamma_{\rm in}^{2}x^{2}}{{\cal
M}^{4}}(1-x^{2}-2
{\cal M}^{2})+\\\ \\
\displaystyle +x^{2}(\gamma_{\rm in}+2 y)^{2}-4
y^{2}-\frac{x^{2}}{{\cal M}^{4}}(1-x^{2}-{\cal
M}^{2})^{2},\label{chisl-1}
\end{array}
\end{equation}
\begin{equation}
\begin{array}{l}
\displaystyle (\gamma_{\rm in}^{2}+x^{2}-1)\frac{d{\cal M}^{2}}{d
x} =2x{\cal M}^{2}-
\frac{\gamma_{\rm in}^{2}x{\cal M}^{2}}{(1-x^{2}-{\cal M}^{2})}+\\\ \\
\displaystyle +\frac{4y^{2}{\cal M}^{6}}{x^{3}(1-x^{2}-{\cal
M}^{2})}. \label{chisl-2}
\end{array}
\end{equation}
We want to emphasize that in the work of \citet{Cyl} these
differential equations were applicable only for the inner part of
the jet, where the integrals of motion could be expressed as the
linear functions of $\Psi$ and $\Omega_{\rm F}=\const$. In fact,
in the case ${\cal M}^{2}\ll x^{2}$ (and that is in the range of
parameters we are interested in) the solution for $B_{\rm p}$ is
the same for an arbitrary function $\Omega_{\rm F}(\Psi)$: the
equation (\ref{chisl-1}) can be rewritten as \citep{Cyl}
\begin{equation}
B_{\rm z}(\varpi)=\frac{4\pi E(\Psi)}{\varpi^{2}\Omega_{\rm
F}^{2}(\Psi)}.
\end{equation}
As $E(\Psi)$ is proportional to the $\Omega_{\rm F}^{2}(\Psi)$,
the particular expression for the angular velocity is not
contained in the equation for $B_{\rm p}$. This ensures the
continuity of the solution even in the region where the condition
$\Omega_{\rm F}=\const$ does not hold.

Analytically, from the set of equations
(\ref{chisl-1})--(\ref{chisl-2}) one can get the following
results:
\begin{enumerate}
\item $x\ll\gamma_{\rm in}$
\begin{equation}
{\cal M}^{2}={\cal M}_{0}^{2}=\const,\; y=\frac{\gamma_{\rm
in}}{2{\cal M}_{0}^{2}}x^{2},
\end{equation}
i.e., the poloidal magnetic field is approximately constant.
\item $x\gg\gamma_{\rm in}$
\begin{enumerate}
\item ${\cal M}_{0}^{2}\gg\gamma_{\rm in}^{2}$
\begin{equation}
{\cal M}^{2}=\frac{{\cal M}_{0}^{2}}{\gamma_{\rm in}^{2}}x^{2},\;
y\propto\ln{\frac{x}{\gamma_{\rm in}}},\label{m-gg-g}
\end{equation}
i.e., the poloidal magnetic field decreases as $B_{\rm p} \propto
\varpi^{-2}$ \citep{clb91,Eichler,bgvl95}.
\item ${\cal M}_{0}^{2}\ll\gamma_{\rm in}^{2}$
\begin{equation}
{\cal M}^{2}=\frac{{\cal M}_{0}^{2}}{\gamma_{\rm in}}x,\;
y=\frac{\gamma_{\rm in}}{2{\cal M}_{0}^{2}} x^2,\label{a-cyl-sol}
\end{equation}
i.e., again $B_{\rm p} \approx$ const.
\end{enumerate}
\end{enumerate}

Using the connection $q=\gamma\mu\eta/E$, one gets for the
Lorentz factor
\begin{equation}
\gamma=\frac{q E}{\mu\eta}=2{\cal M}^{2}\frac{y}{x^{2}}.
\label{gamma-cyl}
\end{equation}
Then, for $x\gg\gamma_{\rm in}$ and ${\cal
M}_{0}^{2}\ll\gamma_{\rm in}^{2}$ the following linear dependence
is valid \citep{Cyl}:
\begin{equation}
\gamma=x.
\end{equation}
Let us find the distance along the axis until which the linear
growth of the Lorentz factor continues. In order to do this one
should write the constant poloidal magnetic field which defines
${\cal M}_{0}^{2}$:
\begin{equation}
B_{\rm z}=\frac{\Psi_{0}\Omega_{\rm F}^{2}}{2\pi\sigma x}\frac{d
y}{d x}=\frac{\Psi_{0}\Omega_{\rm F}^{2}\gamma_{\rm
in}}{2\pi\sigma {\cal M}_{0}^{2}}.
\end{equation}
As this magnetic field should be equal to the outer one, we get
\begin{equation}
z=\sigma\gamma_{\rm in}R_{\rm L},
\end{equation}
and the greatest Lorentz factor near the boundary of the working
volume is
\begin{equation}
\gamma=(\sigma\gamma_{\rm in})^{1/2}.
\end{equation}

For $z > \sigma \gamma_{\rm in} R_{\rm L}$ we shall perform
numerical calculation in the intermediate region between two
limits ${\cal M}_{0}\ll\gamma_{\rm in}$ and ${\cal
M}_{0}\gg\gamma_{\rm in}$.
The boundary condition is the equality of the magnetic flux and
of magnetic field. Thus, for every $z$ we need to find the proper
value of ${\cal M}_{0}^{2}$ that would allow us to make internal
poloidal magnetic field on the border of the working volume be
equal to the external paraboloidal magnetic filed, as the internal
flux becomes equal to the total flux of a jet $\Psi_{\rm jet}$.
In this case the border of the working volume $x_{\rm jet}$ is
defined during the numerical integration. We expect it to be
\begin{equation}
x_{\rm jet}(z)=z^{1/2}.
\end{equation}
The integration shows that it remains almost paraboloidal (see
Fig.~3). Such an integration gives the linear growth of Lorentz
factor until about $\sigma$ (see Fig.~4 and Appendix).

\begin{figure}
\includegraphics{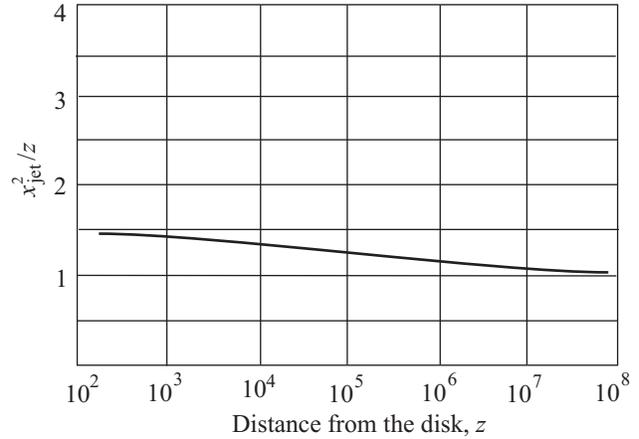}
\caption{The border of the working volume $x_{\rm jet}$ as a
function of $z$ for $\gamma_{\rm in}=8$ and $\sigma=10^{3}$. Its
form remains almost paraboloidal. All the distances are given in
the units of the light cylinder.}
\end{figure}

\begin{figure}
\includegraphics{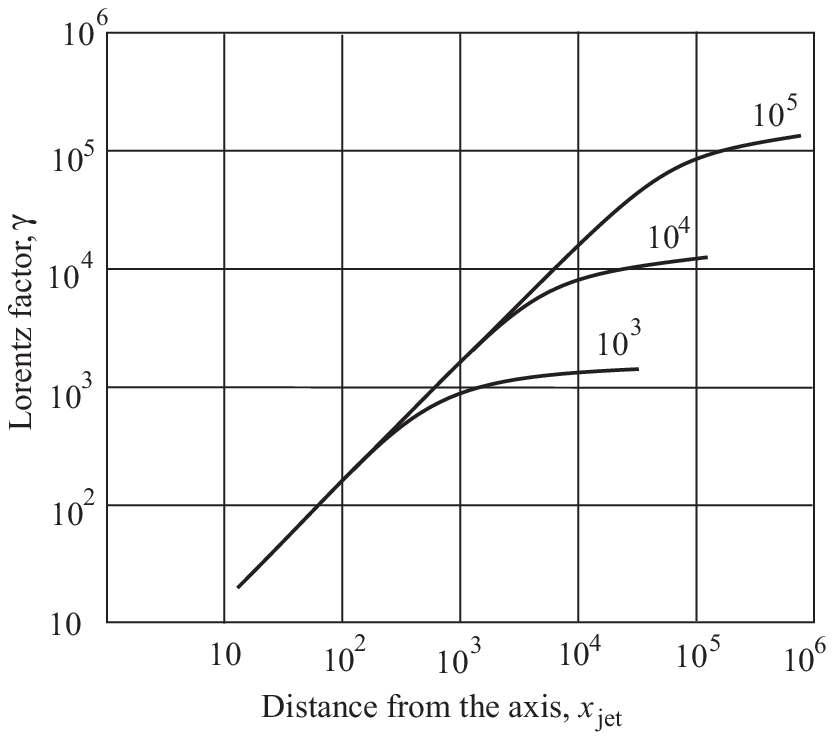}
\caption{The growth of the Lorentz factor $\gamma$ near the
border of the working volume as a function of the distance from
the rotational axis (which is given in the units of the light
cylinder) for different values of $\sigma$ ($10^{3}$, $10^{4}$,
$10^{5}$) in a logarithmic scale. The linear growth of Lorentz
factor continues until $\gamma\approx\sigma$.}
\end{figure}




Thus, the Lorentz factor $\gamma$ grows linearly with the
distance from the axis reaching the value $\sigma$ near the
border of the working volume for $z = \sigma^{2}R_{\rm L}$. This
corresponds to the transformation of about a half of
electro-magnetic energy into the kinetic one. The maximal Lorentz
factor in this problem is $2\sigma$:
\begin{equation}
\frac{E}{\mu\eta}=\gamma_{\rm in}+2\sigma,
\end{equation}
which follows from the definition (\ref{gamma}). Thus, in the
paraboloidal magnetic field the effective particle acceleration
can be realized.

One can see that the condition for the particle acceleration from
the work \citep{Vlah}
\begin{equation}
\frac{\partial}{\partial\Psi}\left(\varpi^{2}B_{\rm P}\right)<0
\end{equation}
holds for our geometry. Indeed, in our case this condition
written for the border of the working volume $\Psi=\Psi_{\rm
jet}$ transforms into the
\begin{equation}
\frac{\partial}{\partial\Psi}\left(\frac{x^{2}_{\rm
jet}(z)}{z}\right)<0.\label{Vlah-cond}
\end{equation}
As the function $x^{2}_{\rm jet}(z)/z$ decreases along the field
line (see Fig.~3), the condition (\ref{Vlah-cond}) is fulfilled.

Here we should point out the main difference of our problem from
the work in which the Michel's monopole solution was taken as the
first approximation.
\begin{enumerate}
\item Even the force-free flow is well collimated already.
\item The curvature term of the paraboloidal problem does not play
any role in the force balance, which allowed us to treat the flow
as one-dimensional in the supersonic domain.
\item The considered flow is supported from outside by the much greater
flux than the one which is contained in the inner region. This
allows the flow to widen and, consequently, to accelerate
outflowing particles inside the working volume.
\end{enumerate}


\section{More realistic models}

In the previous section we were discussing a model in which the
rotational velocity $\Omega_{\rm F}$ of the field lines, which
were assumed to penetrate a horizon of a black hole, in the
working volume was constant, and it was equal to zero outside the
working volume. This allowed us to regard the vacuum field
outside the working volume as confining the flow. In this case
the particle are accelerated effectively up to $\sigma$. However,
such model does not seem to be realistic since we assume the
non-rotating disk.

In this section we shall discuss two models. They would have
three distinct regions. The first one is the working volume with
the given law for rotational velocity, where we assume the MHD
flow of the electron-positron plasma produced by the
Blandford-Znajek process. The second one is the region of the,
presumably, slow ion wind originating from the disk rotating with
Keplerian velocity
\begin{equation}
\Omega_{\rm F}(y)=\Omega_{\rm
F}(\sigma)\left(\frac{\sigma}{y}\right)^{3/2},
\end{equation}
where we choose the parameters so that the function $\Omega_{\rm F
}(y)$ would be continuous on the border of the working volume. The
third region is the vacuum field where the disk angular velocity
drops almost to zero. For our convenience we put $\Omega_{\rm F}$
being equal to zero at the distance where the Keplerian velocity
drops ten times less than the rotational velocity near the axis.
So we still have some portion of the vacuum magnetic flux to
support our flow configuration.

The first model would be still characterized by $\Omega_{\rm
F}=\const$ inside the working volume. For the second one we
employ the following rotational velocity $\Omega_{\rm F}$,
operating in the working volume:
\begin{equation}
\Omega_{\rm F}=\frac{(1-x^{2})[1+\ln(1+x)]}
{\left\{4\ln{2}+1-x^{2}+[1-x^{2}-2(1+x)]\ln(1+x)\right\}}.
\end{equation}
Here $x=\cos{\theta_{\rm H}}$, and $\theta_{\rm H}$ is the
spherical angle at the horizon labeling the field line. It was
first found in the work of \citet{BlandZnajek} for the
paraboloidal magnetic field threading the slowly rotating black
hole. So the second model must be the most realistic.


For these two models we would perform the numerical calculation
for the following set of the ordinal differential equations:
\begin{equation}
\begin{array}{l}
\displaystyle A^{2}\left(\frac{d y}{d x}\right)^{2}=
\frac{\gamma_{\rm in}^{2}x^{2}}{{\cal M}^{4}}(A-
{\cal M}^{2})+\\\ \\
\displaystyle +x^{2}\left(\gamma_{\rm in}+2 \frac{\Omega_{\rm
F}^{2}(y)}{\Omega_{\rm F}^{2}(0)} y\right)^{2}-4 \frac{\Omega_{\rm
F}^{2}(y)}{\Omega_{\rm F}^{2}(0)} y^{2}-\frac{x^{2} A^{2}}{{\cal
M}^{4}},\label{chisl-omega-1}
\end{array}
\end{equation}
\begin{equation}
\begin{array}{l}
\displaystyle \left(\gamma_{\rm in}^{2}+\frac{\Omega_{\rm
F}^{2}(y)}{\Omega_{\rm F}^{2}(0)}x^{2}-1\right)\frac{d{\cal
M}^{2}}{d x} = \frac{4 y^{2} {\cal M}^{6}}{A
x^{3}}\frac{\Omega_{\rm F}^{2}(y)}{\Omega_{\rm F}^{2}(0)}-\\\ \\
\displaystyle -\frac{x {\cal M}^{2}}{A}\frac{\Omega_{\rm
F}^{2}(y)}{\Omega_{\rm F}^{2}(0)}(\gamma_{\rm
in}^{2}-2A)+\frac{{\cal M}^{2}}{2}\frac{d y}{d
x}\left(\frac{x^{2}}{\Omega_{\rm F}^{2}(0)}\frac{d \Omega_{\rm F
}^{2}(y)}{d y}\right). \label{chisl-omega-2}
\end{array}
\end{equation}
The boundary condition again is the equality of the magnetic flux
and of magnetic field at the border between the vacuum field and
the internal flow with the variable rotational velocity
$\Omega_{\rm F}$. We shall be interested in the characteristics
of the flow in the working volume, assuming that the wind from
the disk is governed by the MHD equations.


For the first model we see the same law for the particle Lorentz
factor as a function of the $x$ on the border of the working
volume as the one, we have gotten in the previous section. But now
the border is compressed greater in comparison with its shape in
the previous section (see Fig.~5), so the growth of the Lorentz
factor along the axis is slightly slower.

\begin{figure}
\includegraphics{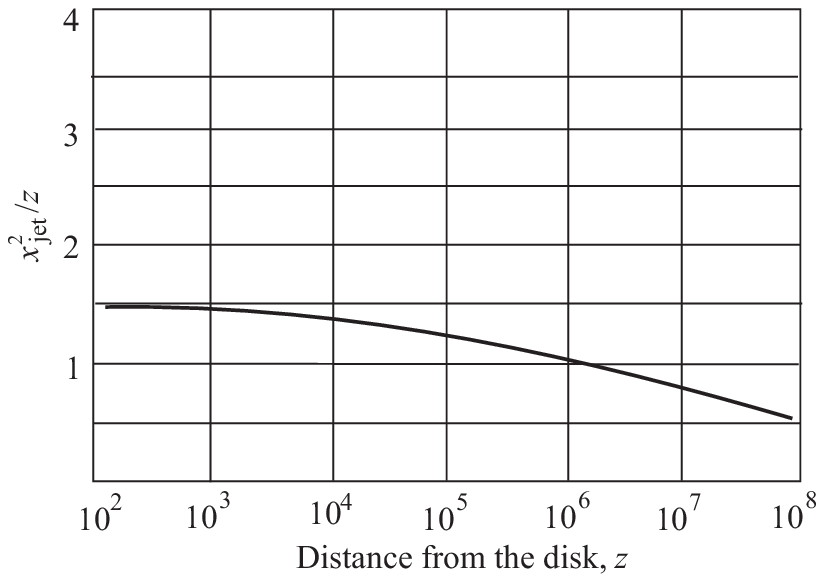}
\caption{The border of the working volume $x_{\rm jet}$ as a
function of $z$ for $\gamma_{\rm in}=8$ and $\sigma=10^{3}$. This
curve is the same for both the first and the second models. Its
form is only slightly compressed in comparison with parabola. All
the distances are given in the units of the light cylinder.}
\end{figure}


The first question with the second model is whether the flow in
subsonic region is greatly affected by the non-constant angular
velocity of the field lines. It is easy to show, that the
linearized equation (\ref{r<rf}) for the given function
$\Omega_{\rm F}$ can be written as
\begin{equation}
\begin{array}{l}
\displaystyle \varepsilon X^2\frac{\partial^{2} f}{\partial
X^{2}}+\varepsilon X \frac{\partial f}{\partial
X}\left(1+\frac{5}{2}M\right) -\varepsilon f +\varepsilon
XY\frac{\partial^{2} f}{\partial Y^{2}} + \varepsilon
X\frac{\partial f}{\partial
Y}\\\ \\
\displaystyle + \pi {\cal
C}\left[2qX\left(1+\frac{1}{4}M\right)+X^2\frac{\partial
q}{\partial X}\right]+\frac{4\pi^{2}e'}{\Omega_{\rm
F}^{2}}\left(1-\frac{1}{2}M\right)=0, \label{lin-var-omega}
\end{array}
\end{equation}
where $\displaystyle M=\frac{\Psi}{\Omega_{\rm F}^{2}}\frac{d
\Omega_{\rm F}^{2}}{d\Psi}$ is much smaller than the unity for
$\Omega_{\rm F}X\ll 1$.

\begin{figure}
\includegraphics{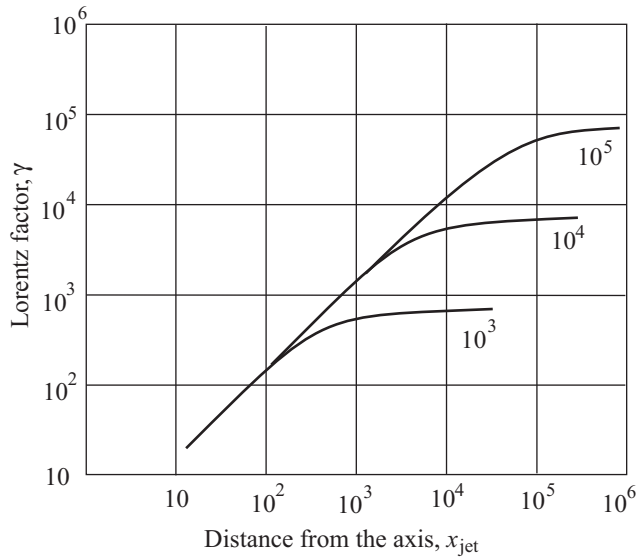}
\caption{The growth of the Lorentz factor $\gamma$ near the
border of the working volume as a function of the distance from
the rotational axis (which is given in the units of the light
cylinder) for different values of $\sigma$ ($10^{3}$, $10^{4}$,
$10^{5}$) in a logarithmic scale is presented for the second
model. Again the linear growth of Lorentz factor takes place, but
it grows until approximately $0.5\sigma$.}
\end{figure}

The numerical integration shows again that the border of the
working volume is compressed greater than it was for the vacuum
field just outside the working volume. Besides, the growth of the
Lorentz factor, although sustaining the shape of the previous
curve $\ln{\gamma}(\ln{x})$, stops at smaller values than in the
case of a constant rotational velocity (see Fig.~6). this results
from the diminishing of the $\Omega_{\rm F}$ near the border of
the working volume:
\begin{equation}
\gamma_{\rm max}=2\sigma\frac{\Omega_{\rm
F}^{2}(\sigma)}{\Omega_{\rm F}^{2}(0)}.
\end{equation}
As the $\Omega_{\rm F}^{2}(\sigma)/\Omega_{\rm F}^{2}(0)\approx
0.28$ for the Blandford-Znajek solution, so does the Lorentz
factor at the border of the working volume. We may conclude that
the decrease of the function $\Omega_{\rm F}$ generally
suppresses the effectiveness of particle acceleration.


\section{Discussion and Astrophysical Applications}

In reality, there are at any way two reasons why the ideal
picture under consideration may be destroyed. First of all, for
large enough $\gamma$ the drag force $F_{\rm drag} \propto
\gamma^2$ can be important. As was demonstrated in \citet{BZS},
for monopole magnetic field it takes place when the compactness
parameter $l_{\rm A}$ becomes larger than $\sigma^{1/3}$. Second,
the particle acceleration continues only until the paraboloidal
poloidal magnetic field can be realized. This can be stopped by
the external magnetic field, say, by the galactic chaotic
magnetic field $B_{\rm gal}\sim 10^{-6}$ G. The flow inside the
border $\Omega_{\rm F}X=\const$ widens, while the external
paraboloidal magnetic field $B_{\rm ext}$ is greater than the
$B_{\rm gal}$. When they equate, the expansion stops and the
particle acceleration ceases. The distance $x_{\rm st}$ from the
axis at which $B_{\rm gal}=B_{\rm ext}$ is given by $x_{\rm
st}^{2}=B_{0}/B_{\rm ext}$. Here $B_{0} = \Psi_{\rm
jet}\Omega_{\rm F}^2$ is the magnetic field strength in the
vicinity of the central object. Thus,
\begin{equation}
\gamma_{\rm max}=\Min\{x_{\rm st},\sigma\}. \label{gamma-max}
\end{equation}
The relation
\begin{equation}
\gamma \approx \left(\frac{B_0}{B_{\rm ext}}\right)^{1/2}
\end{equation}
was already obtained in \citet{Cyl}. As can be easily seen from
(\ref{gamma-max}), whether the Poynting flux will be transformed
into the particle kinetic energy flux depends on the value of
$\sigma$ and $B_{\rm ext}$.

We now consider several astrophysical applications.

\subsection{Active Galactic Nuclei}

For AGNs the central engine is assumed to be a rotating black
hole with mass $M \sim 10^9M_{\odot}$, $R \sim 10^{14}$ cm, the
total luminosity $L \sim 10^{45}$ erg s$^{-1}$, $B_0 \sim 10^4$ G.
In the Michel magnetization parameter $\sigma$ (\ref{sigma})
\begin{equation}
\sigma \approx 10^{14}\lambda^{-1}M_9B_4\left(\frac{\Omega
R}{c}\right),
\end{equation}
the main uncertainty comes from the multiplication parameter
$\lambda$, i.e., in the particle number density $n$. Indeed, for
an electron-positron outflow this value depends on the efficiency
of pair creation in the magnetosphere of a black hole, which is
still undetermined. In particular, this process depends on the
density and energies of the photons in the immediate vicinity of
the black hole. As a result, if the hard-photon density is not
high, then the multiplication parameter is small ($\lambda \sim
10 - 100$; \citet{BeskIstPar,HirOk}). In this case for $(\Omega
R/c) \sim 0.1$--$0.01$ we have $\sigma \sim 10^9 - 10^{12}$.
Knowing $\sigma$ we can estimate the maximal Lorentz factor
$\gamma_{\rm max}$ (\ref{gamma-max}). In the presence of the
external magnetic field of typical value $B_{\rm ext}=10^{-6}$G,
$\gamma_{\rm max}=10^{5}\ll\sigma$, so only the small part of the
Poynting flux can be transformed into the particle flux. On the
other hand, if the density of photons with energies ${\cal
E}_{\gamma}
> 1$MeV is high enough, direct particle creation $\gamma + \gamma
\rightarrow e^+ +e^-$ results in an increase of the particle
density \citep{Sven}. This gives $\sigma \sim 10 - 10^3$. In this
case the Lorentz factor is $\gamma_{\rm max}=\sigma$, and the
energy transformation can be efficient.

Recently the work of \citet{Gracia} has been published in the
electronic arXiv with an idea close to the one presented here
(the central relativistic flow with the disk wind surrounding
it), although with different formalization. In that work the
comparison of the observational data for the jet opening angle
with the model opening angle is presented. Here we can repeat the
same for our paraboloidal inner flow. The distance for the angle
range from the work of \citet{Biretta} lies well inside a sphere
with the radius $z=R_{\rm L}\gamma_{\rm in}\sigma$, so we shall
use the strictly paraboloidal form of the border of the working
volume. We have one free parameter in this case: the value of
$\Omega_{\rm F0}$ which we chose to be equal to $8.2\cdot
10^{-8}$. The result is presented in the Fig.~7. Although the
analytical curve fails to explain the points at an angle $\approx
33$ degrees, it fits well the outer region and (surprisingly) the
innermost point. The Lorentz factor for such distances is
$\approx \gamma_{\rm in}$, so the jet possesses the same moderate
Lorentz factor favoured in the works of \citet{Biretta95},
\citet{Biretta99} and \citet {Cram}.

\begin{figure}
\includegraphics{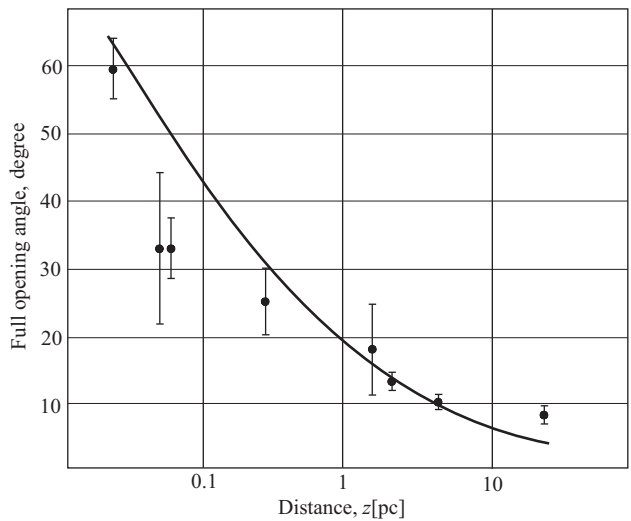}
\caption{The observational data of the opening angle for M87 and
the border of the working volume for $\Omega_{\rm F}(0)=8.2\cdot
10^{-8} \; s^{-1}$.}
\end{figure}

\subsection{Radio Pulsars}
For radio pulsars the central engine is a rotating neutron star
with $M \sim M_{\odot}$, $R \sim 10^{6}$ cm,
and $B_0 \sim 10^{12}$ G. In this case the magnetization parameter
$\sigma \sim 10^4$--$10^6$, corresponding to relativistic
electron-positron plasma, is known with rather high accuracy
(see, e.g., \citet{bgvl2}). The Lorentz factor in the presence of
an external magnetic field is $\gamma_{\rm max}=10^{4}\sim\sigma$.

This estimate is valid, of course, only for the aligned rotator.
In a case of an orthogonal rotator the situation with
$\sigma$-problem is more explicit, as in this case the
Goldriech-Julian charge density must be the $(\Omega R/c)^{1/2}$
smaller near the polar caps than that for the axisymmetric
magnetosphere. Thus, it is natural to expect that the longitudinal
current flowing along the open field lines would be proportionally
smaller too \citep{lsurf1}. But, consequently, toroidal magnetic
field will be smaller than the poloidal electric field in the
vicinity of the light cylinder. On the other hand, it is known
for the Michel's monopole solution that in order to remove the
light surface to infinity, the toroidal magnetic field must be of
the same order as the poloidal electric field on the light
cylinder. If the longitudinal current $j$ does not exceed by
$(\Omega R/c)^{-1/2}$ times the quantity $\rho_{\rm GJ}^{90}c$,
where $\rho_{\rm GJ}^{90}$ is an average charge density on the
polar cap with $\chi \sim 90^{\circ}$ (for the typical pulsars
this factor approach the value of $10^2$), the light surface for
the orthogonal rotator must be located in the vicinity of the
light cylinder. In this case the effective energy conversion and
the current closure takes place in the boundary layer near the
light surface \citep{lsurf1,lsurf2,lsurf3}.

\subsection{Cosmological Gamma-Ray Bursts}

For cosmological gamma-ray bursts the central engine is
represented by the merger of very rapidly orbiting neutron stars
or black holes with $M \sim M_{\odot}$, $R \sim 10^{6}$ cm, and
total luminosity $L \sim 10^{52}$ erg s$^{-1}$ (see, e.g.,
\citet{Lee} for detail). On the other hand, even for a
superstrong magnetic field of $B_0 \sim 10^{15}$ G (which is
necessary to explain the total energy release) the magnetization
parameter $\sigma$ is small ($\sigma < 1 - 100$), because within
this model the magnetic field itself is secondary and its energy
density cannot exceed the plasma energy density. Would it be not
so, i.e., the magnetic field would be prior to the particle
energy flow, we could formally apply our estimate. Having the
external magnetic field to be the order of $10^{11}$ G, we can
get the standard value for the Lorentz factor $\gamma_{\rm max}
\sim 10^{2}$. However, in this case it is hard to explain of the
value of $\sigma$ being the order of $10^{2}$.


\section{Conclusion}

We have gotten the characteristics of the flow in the paraboloidal
magnetic field within the approach of stationary axisymmetric
magnetohydrodynamics. To simplify the problem, we assumed that in
the strongly magnetized flow with $\sigma\gg 1$ a particle inertia
could be described as a small disturbance to the force-free flow.
As the zero approximation the solution with paraboloidal magnetic
field \citet{Bland} was taken.

The position of the fast magnetosonic surface is found to be
$r_{\rm F} \approx (\sigma/\theta)^{1/2}R_{\rm L}$, with the
Lorentz factor changing from $\gamma_{\rm in}$ to $\sigma^{1/3}$
on it. The disturbance $\varepsilon f$ to the stream function
$\Psi$ is $\varepsilon f=\pi{\cal C}
(\Omega_{F}^{2}XY)^{1/2}/\sigma\ll\Psi_{0}$ inside the FMS. As to
the Lorentz factor $\gamma$, it grows linearly with the distance
from the axis.

On the fast magnetosonic surface the structure of the flow may
change significantly. It is implicitly confirmed by the fact that
the characteristics of the flow, which we got under the assumption
of the small disturbance in the supersonic region, is not in the
agreement with the results on the FMS. However, in our problem
the curvature term does not play role in the force balance on the
magnetic surface, and the positions of the FMS in the cylindrical
and paraboloidal flows coincide. These facts allowed us to regard
the problem as one-dimensional and to perform numerical
calculations.

As a result, we got the further growth $\gamma=(z/R_{\rm
L})^{1/2}$ of the Lorentz factor until it reaches the value of
$\sigma$ near the border of the working volume for $z \sim
\sigma^2 R_{\rm L}$ from the equatorial plane. This corresponds
to almost the full conversion of the Poynting energy flux into
the particle kinetic one.

For the more realistic model with three regions --- the jet with
the Blandford-Znajek rotational velocity, the disk wind, and the
vacuum field --- the particle acceleration is suppressed, but
still the maximal Lorentz factor is of the order of $\sigma$.

We want to emphasize that our solution cannot be described in a
self-similar way \citep{Cont,Vlah} as we assumed the constant
angular velocity of the magnetic surfaces near the rotational
axis. This structure cannot be considered within the self-similar
approach.

\section*{Acknowledgments}

We thank R.D.Blandford for the suggestion to use the non-constant
angular velocity for the slowly rotating black hole and M.Sikora
for the proposal to include the rotating disk in the model. We
also thank A.V.Gurevich for his interest and support, H.K.Lee, and
K.A.Postnov for useful discussions. This work was partially
supported by the Russian Foundation for Basic Research (Grant
no.~05-02-17700) and Dynasty fund.

\appendix

\section[]{The analytical solution for the one-dimensional flow}

Let us consider the set of equations
(\ref{chisl-1})--(\ref{chisl-2}). For $x\gg\gamma_{\rm in}$ and
$y\gg\gamma_{\rm in}$ the system can be rewritten as
\begin{equation}
(x^{2}+{\cal M}^{2})^{2}\left(\frac{dy}{dx}\right)^{2}=4
x^{2}y^{2},\label{sys-1}
\end{equation}
\begin{equation}
x^{2}\frac{d{\cal M}^{2}}{dx}=2x{\cal M}^{2} +\frac{4y^{2}{\cal
M}^{6}}{x^{3}(1-x^{2}-{\cal M}^{2})}.\label{sys-2}
\end{equation}
Making the following substitutions
\begin{equation}
q=\frac{{\cal M}^{2}}{x^{2}},\; t=\frac{1}{x^{2}},\; z=y^{2},
\end{equation}
and treating $z$ as a new variable, one can get the following set:
\begin{equation}
\frac{d q}{d z}=-t q^{3},
\end{equation}
\begin{equation}
\frac{d t}{d z}=-\frac{t(1+q)}{2z}.
\end{equation}
It is convenient to transform this set into the one second-order
differential equation
\begin{equation}
\frac{d^{2}q}{d z^{2}}+\frac{1+q}{2 z}\cdot\frac{d q}{d
z}-\frac{3}{q}\cdot\left(\frac{d q}{d z}\right)^{2}=0.\label{ode}
\end{equation}
Knowing its solution, one can get the function $t(z)$ as
\begin{equation}
t(z)=-\frac{d q/d z}{q^{3}}.\label{ode2}
\end{equation}
The solution of (\ref{ode}) gives
\begin{equation}
x=\frac{2a\sqrt{b(q-q_{1})(q-q_{2})}}{q}\left|\frac{q-q_{1}}{q-q_{2}}\right|^{1/\sqrt{1-b}},
\label{sol-1}
\end{equation}
\begin{equation}
y=\frac{ab(q-q_{1})(q-q_{2})}{q^{2}}\left|\frac{q-q_{1}}{q-q_{2}}\right|^{1/\sqrt{1-b}},
\label{sol-2}
\end{equation}
\begin{equation}
\gamma=\frac{\sqrt{b}}{2}\cdot\frac{\sqrt{(q-q_{1})(q-q_{2})}}{q+1}x.
\end{equation} Here
\begin{equation}
q_{1}=\frac{1-\sqrt{1-b}}{b},\; q_{2}=\frac{1+\sqrt{1-b}}{b},
\end{equation}
$a$ and $b$ are constant, $0<a$, $-\infty<b<1$. For $b<0$ the
maximal value of $q$ is equal to unity. On the other hand, for
$b>0$ the value of $q$ is not limited. Thus, we would consider
only the case $b>0$. For the ${\cal M}_{0}^{2}<\gamma_{\rm
in}^{2}$ the solution is represented by the lower branch, and the
case ${\cal M}_{0}^{2}>\gamma_{\rm in}^{2}$ by the upper branch
of the graph (see Fig.~A1).
\begin{figure}
\includegraphics{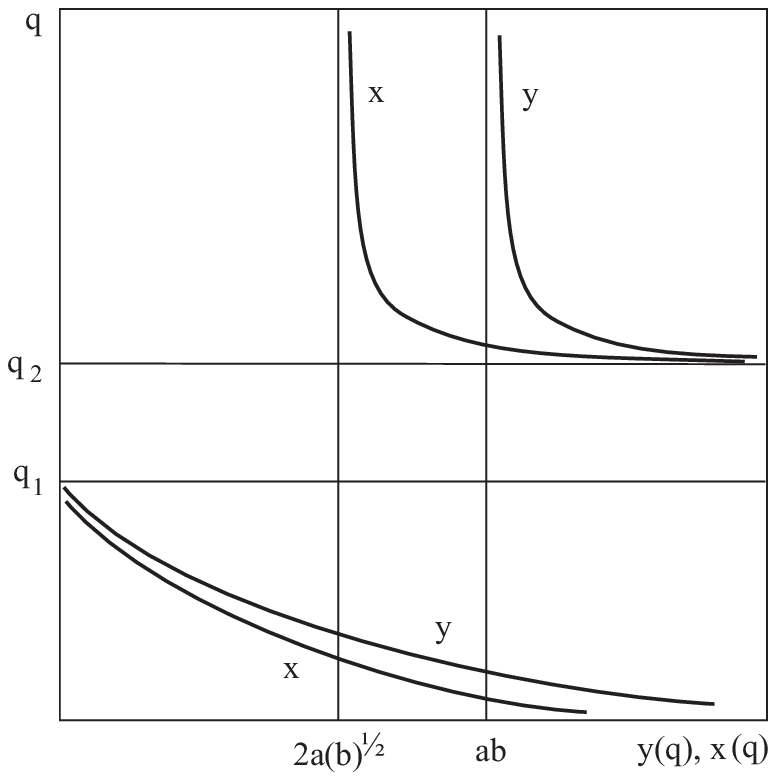}
\caption{The functions x(q) and y(q).}
\end{figure}

Let us find the values of $a$ and $b$. In order to do this we
shall regard two limits of this solution: for ${\cal
M}_{0}^{2}\ll\gamma_{\rm in}^{2}$ and for ${\cal
M}_{0}^{2}\gg\gamma_{\rm in}^{2}$. The condition ${\cal
M}_{0}^{2}\ll\gamma_{\rm in}^{2}$ holds when $q\ll q_{1}<q_{2}$,
so we get the known solution
\begin{equation}
y=A x^{2}=\frac{1}{4
a}\left(\frac{q_{2}}{q_{1}}\right)^{1/\sqrt{1-b}}x^{2},
\end{equation}
\begin{equation}
q=\frac{B}{x}=2 a\sqrt{b
q_{1}q_{2}}\left(\frac{q_{1}}{q_{2}}\right)^{1/\sqrt{1-b}}\frac{1}{x}.
\end{equation}
Equations $A=\gamma_{\rm in}/(2 M_{0}^{2})$ and
$B=M_{0}^{2}/\gamma_{\rm in}$ are not independent and give the
following connection between $a$ and $b$:
\begin{equation}
a=\frac{{\cal M}_{0}^{2}}{2\gamma_{\rm
in}}\left(\frac{1+\sqrt{1-b}}{1-\sqrt{1-b}}\right)^{1/\sqrt{1-b}}.
\end{equation}
When $M_{0}^{2}\gg\gamma_{\rm in}^{2}$
$q=\const=M_{0}^{2}/\gamma_{\rm in}^{2}$, which corresponds to the
asymptotical approach of $q$ to $q_{2}$. Thus, for
$q_{2}=M_{0}^{2}/\gamma_{\rm in}^{2}$, one can get
\begin{equation}
b=\frac{2 {\cal M}_{0}^{2}/\gamma_{\rm in}^{2}-1}{\left({\cal
M}_{0}^{2}/\gamma_{\rm in}^{2}\right)^{2}}.
\end{equation}
Thus,
\begin{equation}
a=\frac{{\cal M}_{0}^{2}}{2\gamma_{\rm in}}\left(2\frac{{\cal
M}_{0}^{2}}{\gamma_{\rm in}^{2}}-1\right)^{\frac{{\cal
M}_{0}^{2}/\gamma_{\rm in}^{2}}{{\cal M}_{0}^{2}/\gamma_{\rm
in}^{2}-1}}.
\end{equation}

Finally, for ${\cal M}_{0}^{2}\ll\gamma_{\rm in}^{2}$ (see the
lower branch in the Fig.~A1) the expressions (\ref{a-cyl-sol})
hold. In the intermediate region ${\cal
M}_{0}^{2}\approx\gamma_{\rm in}^{2}$ (the upper branch in the
Fig.~A1) nothing can be said except the general formulas
(\ref{sol-1},\ref{sol-2}). For ${\cal M}_{0}^{2}\gg\gamma_{\rm
in}^{2}$ (the upper branch in the Fig.~A1, $q\rightarrow q_{2}$)
we can get the refined expressions for $y(x)$ and ${\cal
M}^{2}(x)$. Decomposing the functions
(\ref{sol-1})--(\ref{sol-2}) near $q_{2}$ as $q-q_{2}=\delta$, we
get
\begin{equation}
x=C_{1}\delta^{(\sqrt{1-b}-2)/2\sqrt{1-b}},
\end{equation}
\begin{equation}
y=C_{2}\delta^{(\sqrt{1-b}-1)/\sqrt{1-b}}.
\end{equation}
Thus,
\begin{equation}
y=A x^{2(\sqrt{1-b}-1)/(\sqrt{1-b}-2)}=A
x^{\frac{2}{1+M_{0}^{2}/\gamma_{\rm in}^{2}}},\label{y}
\end{equation}
where $A$ is constant. Thus, for $q=\const$ the expression for $y$
is not $\ln{x}$, but the exponential function of $x$ with the
index depending on the quotient $M_{0}^{2}/\gamma_{\rm in}^{2}$,
which changes from 1 to 0 (compare this result to the numerical
calculation in Fig.~6). Lorentz factor is expressed in this case
by (see Fig.~7)
\begin{equation}
\gamma\propto x^{\frac{2}{1+M_{0}^{2}/\gamma_{\rm
in}^{2}}}.\label{gamma-anal}
\end{equation}
So, if we assume that the effective grow of the Lorentz-factor
continues until the exponent in (\ref{gamma-anal}) is equal to
$1/2$, we can estimate the maximal Lorentz-factor by the value
$\displaystyle\frac{3\sigma}{2}$ (compare it with the Fig.~4).

\bsp

\label{lastpage}


\begin{thebibliography}{99}

\bibitem[\protect\citeauthoryear{Begelman, Blandford \& Rees}{1984}]
{BBR} Begelman~M.C., Blandford~R.D., Rees~M.J.,\ 1984, Rev. Mod.
Phys., 56, 255

\bibitem[\protect\citeauthoryear{Begelman \& Li}{1994}]{acc}
Begelman~M.C., Li~Zh.-Yu., 1994, ApJ, 426, 269

\bibitem[\protect\citeauthoryear{Beskin}{1997}]{Beskin97} Beskin~V.S.,
1997, Phys. Uspekhi, 40, 659

\bibitem[\protect\citeauthoryear{Beskin, Gurevich \&
Istomin}{1993}]{lsurf1} Beskin~V.S., Gurevich~A.V., Istomin~Ya.N.,
1993, Physics of the pulsar magnetosphere, Cambridge University
Press

\bibitem[\protect\citeauthoryear{Beskin, Istomin \& Pariev}{1992}]{BeskIstPar}
Beskin~V.S., Istomin~Ya.N., Pariev~V.I., 1992, Sov. Astron., 36,
642

\bibitem[\protect\citeauthoryear{Beskin, Kuznetsova \& Rafikov}{1998}]{BKR}
Beskin~V.S., Kuznetsova~I.V., Rafikov~R.R., 1998, MNRAS, 299, 341

\bibitem[\protect\citeauthoryear{Beskin \& Malyshkin}{2000}]{Cyl} Beskin~V.S, Malyshkin~L.M.,
2000, Astron. Lett., 26, 208

\bibitem[\protect\citeauthoryear{Beskin \& Okamoto}{2000}]{BeskinOkamoto} Beskin~V.S, Okamoto~I.,
2000, MNRAS, 313, 445

\bibitem[\protect\citeauthoryear{Beskin \&
Rafikov}{2000}]{lsurf3} Beskin~V.S., Rafikov~R.R., 2000, MNRAS,
313, 433

\bibitem[\protect\citeauthoryear{Beskin, Zakamska \& Sol}{2003}]{BZS}
Beskin~V.S., Zakamska~N.L., Sol~H., 2004, MNRAS, 347, 587

\bibitem[\protect\citeauthoryear{Biretta et al.}{1995}]{Biretta95}
Biretta~J.A., Zhou~F., Owen~F.N., 1995, ApJ, 447, 582

\bibitem[\protect\citeauthoryear{Biretta et al.}{1999}]{Biretta99}
Biretta~J.A., Sparks~W.B., Macchetto~F., 1999, ApJ, 520, 621

\bibitem[\protect\citeauthoryear{Biretta et al.}{2002}]{Biretta}
Biretta~J.A., Junor~W., Livio~M., 2002, New Astronomy Review, 46,
239

\bibitem[\protect\citeauthoryear{Blanford}{1976}]{Bland}
Blandford~R.D., 1976, MNRAS, 176, 465

\bibitem[\protect\citeauthoryear{Blanford \& Payne}{1982}]{BPayne}
Blandford~R.D., Payne~D.G., 1982. MNRAS, 199, 883

\bibitem[\protect\citeauthoryear{Blanford \& Znajek}{1977}]{BlandZnajek}
Blandford~R.D., Znajek~R.L., 1977, MNRAS, 179, 433

\bibitem[\protect\citeauthoryear{Bogovalov}{1995}]{bgvl95}
Bogovalov~S.V., 1995, Astron. Lett., 21, 565

\bibitem[\protect\citeauthoryear{Bogovalov}{1997}]{bgvl2}
Bogovalov~S.V., 1997, A\&A, 327, 662

\bibitem[\protect\citeauthoryear{Camenzind}{1986}]{Camenzind} Camenzind~M.,
1986, A\&A, 162, 32

\bibitem[\protect\citeauthoryear{Chiueh, Li \& Begelman}{1991}]{clb91}
Chiueh~T., Li~Zh.-Yu., Begelman~M.C., 1991, ApJ, 377, 462

\bibitem[\protect\citeauthoryear{Chiueh, Li \&
Begelman}{1998}]{lsurf2} Chiueh~T., Li~Zh.-Yu., Begelman~M.C.,
1998, ApJ, 505, 835

\bibitem[\protect\citeauthoryear{Contopoulos}{1995}]{Cont}
Contopoulos~J., 1995, ApJ, 446, 67

\bibitem[\protect\citeauthoryear{Coroniti}{1990}]{recon1}
Coroniti~V.F., 1990, ApJ, 349, 538

\bibitem[\protect\citeauthoryear{Cramphorn et al.}{2004}]{Cram}
Cramphorn~C.K., Sazonov~S.Y., Sunyaev~R.A., 2004, A\&A, 420, 33

\bibitem[\protect\citeauthoryear{Eichler}{1993}]{Eichler}
Eichler D., 1993, ApJ, 419, 111

\bibitem[\protect\citeauthoryear{Gracia, Tsinganos, Bogovalov}{2005}]{Gracia} Gracia~J.,
Tsinganos~K., Bogovalov~V., A\&A accepted

\bibitem[\protect\citeauthoryear{Hirotani \& Okamoto}{1998}]{HirOk}
Hirotani~K., Okamoto~I., 1998, ApJ, 497, 563

\bibitem[\protect\citeauthoryear{Kennel \& Coroniti}{1984}]{KC}
Kennel~C.F., Coroniti~F.V., 1984, ApJ, 283, 694

\bibitem[\protect\citeauthoryear{Kirk \&
Lyubarsky}{2001}]{recon3}Kirk~J.G., Lyubarsky~Y., 2001, J.
Astron. Soc. Australia, 18, 415

\bibitem[\protect\citeauthoryear{Komissarov}{2004}]{Komm} Komissarov~S.S.,
2004, MNRAS, 350, 1431

\bibitem[\protect\citeauthoryear{Lee \& Park}{2004}]{LeePark} Lee~H.K., Park~J.,
2004, Phys.Rev.D, 70, 063001

\bibitem[\protect\citeauthoryear{Lee at al}{2000}]{Lee} Lee~H.K., Wijers~R.A.M.J.,
Brown~G.E., 2000, Phys.Rep., 325, 83

\bibitem[\protect\citeauthoryear{Li, Chiueh \& Begelman}{1992}]{lcb92}
Li Zh.-Yu., Chiueh T., Begelman M.C., 1992, ApJ, 394, 459

\bibitem[\protect\citeauthoryear{Lyubarsky \&
Kirk}{2001}]{recon2} Lyubarsky~Y., Kirk~J.G., 2001, ApJ, 547, 437


\bibitem[\protect\citeauthoryear{Michel}{1969}]{Mchl} Michel~F.
C., 1969, Astrophys.~J., 158, 727

\bibitem[\protect\citeauthoryear{Michel}{1991}]{michbook} Michel~F.,
1991, Theory of neutron star magnetosphere, Chicago University
Press

\bibitem[\protect\citeauthoryear{Sikora et al}{2005}]{SBML}
Sikora~M., Begelman~M.C., Madejski~G.M., Lasota~J.-P.,
astro-ph/0502115

\bibitem[\protect\citeauthoryear{Spitkovsky \& Arons}{2004}]{SpitArons}
Spitkovsky~A., Arons~J., 2004, ApJ, 603, 669

\bibitem[\protect\citeauthoryear{Svensson}{1984}]{Sven}
Svensson~R., 1984, MNRAS, 209, 175

\bibitem[\protect\citeauthoryear{Thomson, Chang, \& Quataert}{2004}]{Thom}
Thomson~T.A., Chang~P., Quataert~E., 2004, ApJ, 611, 380

\bibitem[\protect\citeauthoryear{Vlahakis}{2004}]{Vlah} Vlahakis~N.,
2004, ApJ, 600, 324

\end{thebibliography}
\end{document}